\documentclass[prb,twocolumn,superscriptaddress,showpacs]{revtex4-1}

\usepackage{graphicx,bm,color,amsmath}

\newcommand{\bfB}{{\bf B}}
\newcommand{\bfq}{{\bf q}}
\newcommand{\bfr}{{\bf r}}
\newcommand{\bfk}{{\bf k}}
\newcommand{\ua}{\uparrow}
\newcommand{\da}{\downarrow}
\newcommand{\dy}{\displaystyle}
\newcommand{\hxc}{h^{\rm xc}}
\newcommand{\exc}{e_{\rm xc}}

\newcommand{\sws}{S_{\rm sw}} % spin-wave stiffness
\begin{document}

\title{Spin precession and  spin waves in a chiral electron gas: beyond Larmor's theorem}

\author{Shahrzad Karimi}
\affiliation{Department of Physics and Astronomy, University of Missouri, Columbia, Missouri 65211, USA}

\author{Florent Baboux}
\affiliation{Institut des Nanosciences de Paris, CNRS/Universit\'e Paris VI, Paris 75005, France}
\affiliation{Laboratoire Mat\'eriaux et Ph\'enom\`enes Quantiques, Universit\'e Paris Diderot, CNRS-UMR 7162, Paris 75013, France}

\author{Florent Perez}
\affiliation{Institut des Nanosciences de Paris, CNRS/Universit\'e Paris VI, Paris 75005, France}

\author{Carsten A. Ullrich}
\affiliation{Department of Physics and Astronomy, University of Missouri, Columbia, Missouri 65211, USA}

\author{Grzegorz Karczewski}
\affiliation{Institute of Physics, Polish Academy of Sciences, Warsaw, Poland}
\author{Tomasz Wojtowicz}
\affiliation{Institute of Physics, Polish Academy of Sciences, Warsaw, Poland}

\date{\today }

\begin{abstract}
Larmor's theorem holds for magnetic systems that are invariant under spin rotation. In the presence of spin-orbit coupling
this invariance is lost and  Larmor's theorem is broken:  for systems of interacting electrons, this gives rise to a subtle interplay
between the spin-orbit coupling acting on individual single-particle states and Coulomb many-body effects.
We consider a quasi-two-dimensional, partially spin-polarized
electron gas in a semiconductor quantum well in the presence of Rashba and Dresselhaus spin-orbit coupling.
Using a linear-response approach based on time-dependent density-functional theory, we calculate the
dispersions of spin-flip waves. We obtain analytic results for small wave vectors and up to second order in the Rashba and Dresselhaus coupling strengths
$\alpha$ and $\beta$. Comparison with experimental data from inelastic light scattering allows us to extract
$\alpha$ and $\beta$ as well as the spin-wave stiffness very accurately. We find significant deviations from the
local density approximation for spin-dependent electron systems.
\end{abstract}


\maketitle

\section{Introduction}

Larmor's theorem \cite{LandauLifshitz,Lipparini} states that in a system of charges, all with the same charge-mass ratio $q/m$,
moving in a centrally symmetric electrostatic potential and in a sufficiently weak magnetic field $\bf B$, the charges precess
about the direction of the magnetic field with the frequency
\begin{equation} \label{I.1}
\Omega_L = g\frac{qB}{2m}
\end{equation}
(in SI units), where $g$ is the gyromagnetic ratio or g-factor.

In condensed-matter physics, Larmor's theorem applies to the long-wavelength limit of
spin-wave excitations in magnetic systems which are invariant under spin rotation. \cite{Yosida}
In particular, the electrons in a two-dimensional electron gas (2DEG) in the presence of a constant uniform magnetic field
carry out a precessional motion at the single-particle Larmor frequency, despite the presence of Coulomb interactions.

If spin-rotational invariance is broken---for instance, in the presence of spin-orbit coupling (SOC)---Larmor's theorem
is no longer guaranteed to hold,
and there will be corrections to $\Omega_L$. This was experimentally observed over three decades ago for a
2DEG in a GaAs/AlGaAs heterostructure, using electron spin resonance (ESR). \cite{Stein1983} Subsequently,
several theoretical studies addressed the breaking of Larmor's theorem in collective spin excitations in
2DEGs.\cite{Longo1993,Califano2006,Zhang2008,Roldan2010,Krishtopenko2015,Maiti2016}
The corrections to $\Omega_L$ are caused by a subtle interplay between SOC and Coulomb many-body effects,
which poses significant formal and computational challenges; on the other hand, this offers interesting opportunities
for the experimental determination of SOC parameters and the study of many-body interactions.

In this paper, we present a joint experimental and theoretical study of the spin-wave dispersions of
a partially spin-polarized 2DEG in a semiconductor quantum well. The influence of Rashba and Dresselhaus SOC on collective electronic modes
in quantum wells was first theoretically predicted to cause an angular modulation of the intersubband plasmon dispersion.
\cite{Ullrich2002,Ullrich2003} The effect was later experimentally confirmed, \cite{Baboux2012} and then
extended to spin-wave dispersions.\cite{Baboux2013,Baboux2015,Baboux2016,Perez2016}

In the absence of SOC, the real part of the spin-wave dispersion of a paramagnetic 2DEG has the following form for small wave vectors:\cite{Perez2009}
\begin{equation} \label{2}
\hbar \omega_{\rm sw}(\bfq) = Z + \frac{1}{2} S_{\rm sw} q^2 \:,
\end{equation}
where $Z$ is the bare Zeeman energy, and $S_{\rm sw}$ is the spin-wave stiffness, which depends on Coulomb many-body effects (explicit
expressions for $Z$ and $S_{\rm sw}$ will be given in Section \ref{sec:II}).
We recently discovered\cite{Perez2016} that, to first order in the Rashba and Dresselhaus spin-orbit coupling strengths
$\alpha$ and $\beta$, the spin-wave dispersion is unchanged apart from a chiral shift by a constant wave vector $\bfq_0$ (defined in Sec. III)
which depends on $\alpha$, $\beta$ and the angle $\varphi$ between the magnetization direction and the [010] crystalline axis (see Fig.~\ref{fig1}).
In other words, to quadratic order in the wave vector, we find
\begin{equation} \label{I.1a}
\hbar \omega_{\rm sw}^{\rm SO}(\bfq) =  Z + \frac{1}{2} S_{\rm sw} |\bfq + \bfq_0|^2 + {\cal O}(\alpha^2,\beta^2).
\end{equation}
The spin-wave stiffness $S_{\rm sw}$ remains unchanged, to leading order in $\alpha,\beta$.
The physical interpretation is that the spin wave behaves as if it were transformed into a spin-orbit twisted reference frame.
This opens up new possibilities for manipulating spin waves, which may lead to new applications in spintronics.

To account for higher-order SOC effects in the spin-wave dispersion, it is sensible to rewrite Eq. (\ref{I.1a}) in a more
general manner:
\begin{equation} \label{I.2}
\omega_{\rm sw}^{\rm SO}(\bfq) = E_0(\varphi) + E_1(\varphi) q + E_2(\varphi) q^2 \:,
\end{equation}
where the coefficients $E_0$, $E_1$ and $E_2$ depend on the propagation direction $\varphi$ (see Fig.~\ref{fig1}).
From Eq. (\ref{I.1a}), the linear coefficient is given to leading order in SOC by $E_1(\varphi) = S_{\rm sw}\bfq\cdot\bfq_0 /q$, which can be expressed as\cite{Perez2016}
\begin{equation}
E_1(\varphi) =-\frac{2}{\zeta} \frac{Z}{(Z^* - Z)}(\alpha + \beta \sin 2\varphi) \:,
\label{eq_modulation_E1}
\end{equation}
where $\zeta$ is the spin polarization of the 2DEG, and $Z^*$ is the renormalized Zeeman splitting, to be defined below in Section IIB.

We will present a linear-response approach based on time-dependent density-functional theory (TDDFT)
which allows us to obtain analytical results for $E_0$, to second order in $\alpha,\beta$, and numerical results for $E_1$ and $E_2$
to all orders in SOC.
The breaking of Larmor's theorem is expressed in the coefficient $E_0$, which has $\varphi$-dependent corrections to $Z$.
In Section IV we will obtain the following result to leading order in SOC:
\begin{eqnarray}
 E_0(\varphi) &=& Z+ \frac{2\pi N_s }{Z^* f_T}
\big[ (\alpha^2 + \beta^2)(3f_T+2) \nonumber\\
&&
{}+ 2\alpha\beta\sin(2\varphi) (f_T+2) \big],
\label{eq_modulation_E0}
\end{eqnarray}
where $f_T = (Z-Z^*)/Z^*$.

Our analytical and numerical results will be compared with experimental results, obtained via inelastic light scattering.
By fitting $E_0$, $E_1$ and $E_2$ we are able to extract values for $Z$, $\alpha$ and $\beta$ and present evidence for the $\varphi$ dependence of $E_0$ and $E_2$, which had not been considered in our earlier work.\cite{Perez2016} Comparison to theory shows significant deviations from the standard approximation in TDDFT, the adiabatic local-density
approximation (ALDA). This provides new incentives to search for better exchange-correlation functionals for transverse
spin excitations of electronic systems.

This paper is organized as follows. In Section \ref{sec:II} we discuss Larmor's theorem without SOC: first, for completeness, we present a
general proof for interacting many-body systems, and then we discuss Larmor's theorem from a TDDFT perspective. This will lead to a new
constraint for the exchange-correlation kernel of linear-response TDDFT. In Section \ref{sec:III} we consider the electronic
states in a quantum well with SOC and an in-plane magnetic field. Section \ref{sec:IV} contains the derivation of the spin-wave dispersions
from linear-response TDDFT, in the presence of SOC.
In Section \ref{sec:V} we compare our theory with experimental results and discuss our findings. Section \ref{sec:VI} gives our conclusions.

%--------------------------------------------------------------------------------------------------------------

\section{Larmor's theorem} \label{sec:II}

In this section we consider Larmor's theorem in a 2DEG, from a general many-body perspective (the proof given in Sec. \ref{sec:IIA} is
not new\cite{Lipparini} but included here to keep the paper self-contained), and from the perspective of TDDFT. This will set
the stage for the discussions in the following sections where the effects of SOC are included.

\subsection{Long-wavelength limit of spin waves a 2DEG} \label{sec:IIA}

Let us consider a 2DEG in the presence of a uniform magnetic field ${\bf B} = B \hat e_z$, where $\hat e_z$ is a unit
vector lying in the plane of the 2DEG. The Hamiltonian is
\begin{equation} \label{2.1}
\hat H = \sum_i\left[ \frac{\hat{\bf p}^2_i}{2m} + \frac{Z}{2} \hat \sigma_{z,i} \right]
+ \frac{e^2}{2} \sum_{ij} \frac{1}{|\bfr_i - \bfr_j|} \:.
\end{equation}
Here, $m$ and $e$ are the electron mass and charge, $Z=g \mu_B B$ is the Zeeman energy (the splitting between the
spin-up and spin-down bands), and $\mu_B=|e|\hbar/2m$ is the Bohr magneton. For a 2DEG embedded in a semiconductor, $m$, $e$, and $g$
are replaced by the effective mass, charge and g-factor, $m^*$, $e^*$ and $g^*$, where $g^*$ could be a positive or negative number.

Since the magnetic field is applied in the plane of the 2DEG  (in this section, we assume for simplicity that the 2DEG has zero thickness),
its only effect is on the electron spin
and there is no Landau level quantization. Later on, when we discuss quantum wells of finite width,
we will exclude situations where the magnetic length $l_B= \sqrt{\hbar/|eB|}$ is smaller than the well width,
and hence continue to disregard any orbital angular momentum contributions.

Let us define the spin-wave operator \cite{Perez2009,Perez2011}
\begin{equation}\label{2.2}
\hat S_{+,\bfq} = \frac{1}{2}\sum_i \hat \sigma_{+,i}e^{-i \bfq \cdot \bfr_i} \:,
\end{equation}
where $\hat \sigma_+ = \hat \sigma_x + i \hat \sigma_y$. This operator satisfies the Heisenberg equation of motion
\begin{equation}\label{2.3}
\frac{d}{dt}\hat S_{+,\bfq} =\frac{1}{i\hbar} [\hat S_{+,\bfq},\hat H] = i \omega_{\rm sw}(\bfq) \hat S_{+,\bfq} \:,
\end{equation}
where $\omega_{\rm sw}(\bfq)$ is the spin-wave frequency dispersion of the 2DEG.
We are interested in the special case $\bfq=0$, and abbreviate $\omega_{\rm sw}(\bfq=0)=\omega_{\rm sw,0}$.
The operator $\hat S_{+,0} = \frac{1}{2}\sum_i \hat \sigma_{+,i}$
commutes with the kinetic and electron-electron interaction parts of $\hat H$, and we obtain
\begin{equation}\label{2.4}
[\hat S_{+,0},\hat H]
=
\frac{Z}{4}\sum_i [\hat \sigma_{+,i},\hat\sigma_{z,i}]  \nonumber\\
=
-Z  \hat S_{+,0} \:,
\end{equation}
where we used the standard commutation relations between the Pauli matrices $\hat \sigma_x$, $\hat \sigma_y$ and $\hat \sigma_z$.
Together with Eq. (\ref{2.3}), this yields
\begin{equation}\label{2.5}
 \frac{d}{dt}\hat S_{+,0} = \frac{i}{\hbar} Z  \hat S_{+,0} \:,
\end{equation}
and hence
\begin{equation}\label{2.6}
\hbar \omega_{\rm sw,0} = Z \:.
\end{equation}
Larmor's theorem thus says that the long-wavelength limit of the spin-wave dispersion of a 2DEG
is given by the bare Zeeman energy, regardless of the presence of Coulomb interactions. By comparison with
Eq. (\ref{I.1}) we have $\Omega_L = Z/\hbar$.

\subsection{TDDFT perspective} \label{sec:IIB}

TDDFT is a formally exact approach to calculate excitations in electronic systems.\cite{Gross1985,TDDFTbook}
In the most general case of a magnetic system, TDDFT can be formulated using the spin-density matrix $\underline n$ as basic variable, whose
elements are defined as
\begin{equation}\label{2.7}
n_{\sigma\sigma'}(\bfr,t) = \langle \Psi(t) | \hat \psi_{\sigma'}^\dagger(\bfr)\hat \psi_\sigma(\bfr) | \Psi(t)\rangle \:,
\end{equation}
where $\Psi(t)$ is the time-dependent many-body wave function, and $\hat \psi_\sigma(\bfr),\psi_{\sigma'}^\dagger(\bfr)$
are fermionic field operators for spins $\sigma$ and $\sigma'$, respectively.
The spin-density matrix is diagonal for spatially uniform magnetic fields if the spin quantization axis is along
the direction of the field. However, spin-flip excitations involve the transverse (off-diagonal) spin-density matrix response.

The frequency- and momentum-dependent linear-response equation for a 2DEG has the following form:
\begin{equation}\label{2.8}
n_{\sigma\sigma'}^{(1)}(\bfq,\omega) = \sum_{\tau\tau'} \chi_{\sigma\sigma',\tau \tau'}^{\rm int}(\bfq,\omega)
v_{\tau\tau'}^{(1)}(\bfq,\omega) \:,
\end{equation}
where $v_{\tau'\tau'}^{(1)}(\bfq,\omega)$ is a spin-dependent perturbation, and $\chi_{\sigma\sigma',\tau \tau'}^{\rm int}(\bfq,\omega)$
is the spin-density matrix response function of the interacting many-body system.

The TDDFT counterpart of Eq. (\ref{2.8}) is
\begin{equation}\label{2.9}
n_{\sigma\sigma'}^{(1)}(\bfq,\omega) = \sum_{\tau\tau'} \chi_{\sigma\sigma',\tau \tau'}(\bfq,\omega)
v_{\tau\tau'}^{(1)\rm eff}(\bfq,\omega) \:,
\end{equation}
where $\chi_{\sigma\sigma',\tau \tau'}(\bfq,\omega)$ is the response function of the corresponding noninteracting 2DEG,
and the effective perturbation is
\begin{eqnarray}\label{2.10}
\delta v_{\tau\tau'}^{(1)\rm eff}(\bfq,\omega)
&=&
v_{\tau\tau'}^{(1)}(\bfq,\omega) \\
&+&
\sum_{\lambda\lambda'} \left[ \frac{2\pi}{q} + f^{\rm xc}_{\tau\tau',\lambda\lambda'}(\bfq,\omega)\right]
n_{\lambda\lambda'}^{(1)}(\bfq,\omega). \nonumber
\end{eqnarray}
Here, $f^{\rm xc}_{\tau\tau',\lambda\lambda'}(\bfq,\omega)$ is the exchange-correlation (xc) kernel for the spin-density matrix
response of the 2DEG.

Let us now consider a noninteracting spin-polarized 2DEG with the Kohn-Sham Hamiltonian
\begin{equation} \label{2.11}
\hat h = \sum_i\left[ \frac{\hat{\bf p}^2_i}{2m} + \frac{Z^*}{2} \hat \sigma_{z,i} \right]\:,
\end{equation}
which produces two parabolic, spin-split energy bands $\hbar^2 k^2/2m + \varepsilon_{\ua,\da}$
(spin-up and spin-down are taken with respect to the $z$ axis). In the following let us assume that
$\varepsilon_\ua - \varepsilon_\da>0$, so $\zeta<0$. The renormalized Zeeman energy is therefore given by
\begin{equation}\label{2.12}
Z^* = \varepsilon_\ua - \varepsilon_\da = Z + v_{\rm xc\ua} - v_{\rm xc\da}\:.
\end{equation}
From the xc energy per particle of a spin-polarized 2DEG,\cite{Attaccalite2002}
$e_{\rm xc}(n,\zeta)$ (where $n$ and $\zeta$ are the
density and spin polarization, respectively), the spin-dependent
xc potentials are obtained as
\begin{eqnarray} \label{2.13}
v_{\rm xc \ua} &=&  e_{\rm xc} + n\: \frac{ \partial  e_{\rm xc}}{\partial n}
+ (1- \zeta)\frac{\partial e_{\rm xc}}{\partial \zeta}
\\
v_{\rm xc \da}&=&  e_{\rm xc}+ n\: \frac{ \partial  e_{\rm xc}}{\partial n}
-  (1+ \zeta)
\frac{\partial e_{\rm xc}}{\partial \zeta} \:,
\end{eqnarray}
so the renormalized Zeeman energy is\cite{Perez2007,Perez2009}
\begin{equation} \label{2.14}
Z^* =  Z + 2\frac{\partial  e_{\rm xc}}{\partial \zeta} \:.
\end{equation}
Now let us calculate the collective spin-flip excitations using linear response theory. Since the ground state of the 2DEG
has no transverse spin polarization, the spin-density-matrix response decouples into longitudinal and transverse
channels, and we can write the associated noninteracting response functions as
\begin{eqnarray} \label{2.15}
\underline \chi_L (\bfq,\omega) &=&
\left(\begin{array}{cc} \chi_{\ua\ua,\ua\ua} & \chi_{\ua\ua,\da\da}\\ \chi_{\da\da,\ua\ua} & \chi_{\da\da,\da\da}\end{array}\right)\\
\underline \chi_T (\bfq,\omega) &=&
\left(\begin{array}{cc} \chi_{\ua\da,\ua\da} & \chi_{\ua\da,\da\ua}\\ \chi_{\da\ua,\ua\da} & \chi_{\da\ua,\da\ua}\end{array}\right),
\label{2.16}
\end{eqnarray}
and similar for the interacting case.
The transverse part of the interacting response function is  diagonal, and can be expressed via TDDFT as
\begin{equation}\label{2.17}
\underline \chi_T^{\rm int} (\bfq,\omega)=
\left( \begin{array}{cc}
\frac{ \chi_{\ua\da,\ua\da}}{1 -  \chi_{\ua\da,\ua\da} f^{\rm xc}_{\ua\da,\ua\da}} & 0 \\
0 & \frac{ \chi_{\da\ua,\da\ua}}{1 -  \chi_{\da\ua,\da\ua} f^{\rm xc}_{\da\ua,\da\ua}} \end{array}\right).
\end{equation}
We now consider the case $\bfq=0$, where the spin-flip Lindhard functions have the simple form
\begin{eqnarray} \label{2.18}
\chi_{\ua\da,\ua\da}(0,\omega)  &=&
-\frac{n\zeta}{\omega - Z^*}\\
\chi_{\da\ua,\da\ua}(0,\omega)  &=&
\frac{n\zeta}{\omega + Z^*}
\label{2.19}
\end{eqnarray}
(for a comprehensive discussion of the Lindhard func\-tion---the response function of the noninteracting electron gas---see
Ref. \onlinecite{GiulianiVignale}).
We get a collective excitation at that frequency where $\underline \chi_T^{\rm int}$ is singular. We substitute
Eqs. (\ref{2.18}) and (\ref{2.19}) into Eq. (\ref{2.17}) and set the determinant of the $2\times 2$ transverse
response matrix $\underline\chi_T^{\rm int}$ to zero. Furthermore,
because the system has no transverse spin polarization in the ground state, we have
\begin{equation}\label{2.20}
f^{\rm xc}_{\ua\da,\ua\da}(q,\omega) = f^{\rm xc}_{\da\ua,\da\ua}(q,\omega)
\equiv f^{\rm xc}_T(q,\omega)\:.
\end{equation}
This yields the $\bfq=0$ limit of the spin-flip wave of the 2DEG as
\begin{equation}\label{2.21}
\omega_{\rm sw,0} = Z^* - n\zeta f^{\rm xc}_T(0,\omega_{\rm sw,0}).
\end{equation}
This expression is formally exact. Comparing with the many-body result (\ref{2.6}), and using
Eq. (\ref{2.14}), gives
\begin{equation}\label{2.22}
f^{\rm xc}_T(0,Z) =  \frac{2}{n\zeta} \frac{\partial e_{\rm xc}}{\partial \zeta} \:.
\end{equation}
%We can replace the bare Zeeman energy $Z$ in the argument of $f^{\rm xc}_T$ by noting that
%$\zeta = -Z^*/2\pi n$, and therefore
%\begin{equation}\label{2.23}
%Z = -2\pi n \zeta - 2 \frac{\partial e_{\rm xc}}{\partial \zeta} \:.
%\end{equation}
Equation (\ref{2.22}) is an exact constraint on the transverse xc kernel of the 2DEG,
based on Larmor's theorem. It is not difficult to show that it is satisfied by the adiabatic local-density approximation (ALDA),
where the xc kernel is frequency- and momentum-independent.\cite{Rajagopal1978,Perez2009}

For small but finite wave vectors, one obtains the long-wavelength spin-flip wave dispersion:\cite{Perez2009}
\begin{equation}\label{2.24}
    \omega_{\rm sw,q}=Z-\frac{1}{|\zeta|}\frac{Z}{Z^\ast-Z}\frac{\hbar}{2m^\ast}q^2
\end{equation}
which yields the spin-wave stiffness $S_{\rm sw}=-\frac{1}{|\zeta|}\frac{Z}{Z^\ast-Z}\frac{\hbar}{m^\ast}$,
see Eq. (\ref{2}).

\begin{figure}
\includegraphics[width=\linewidth]{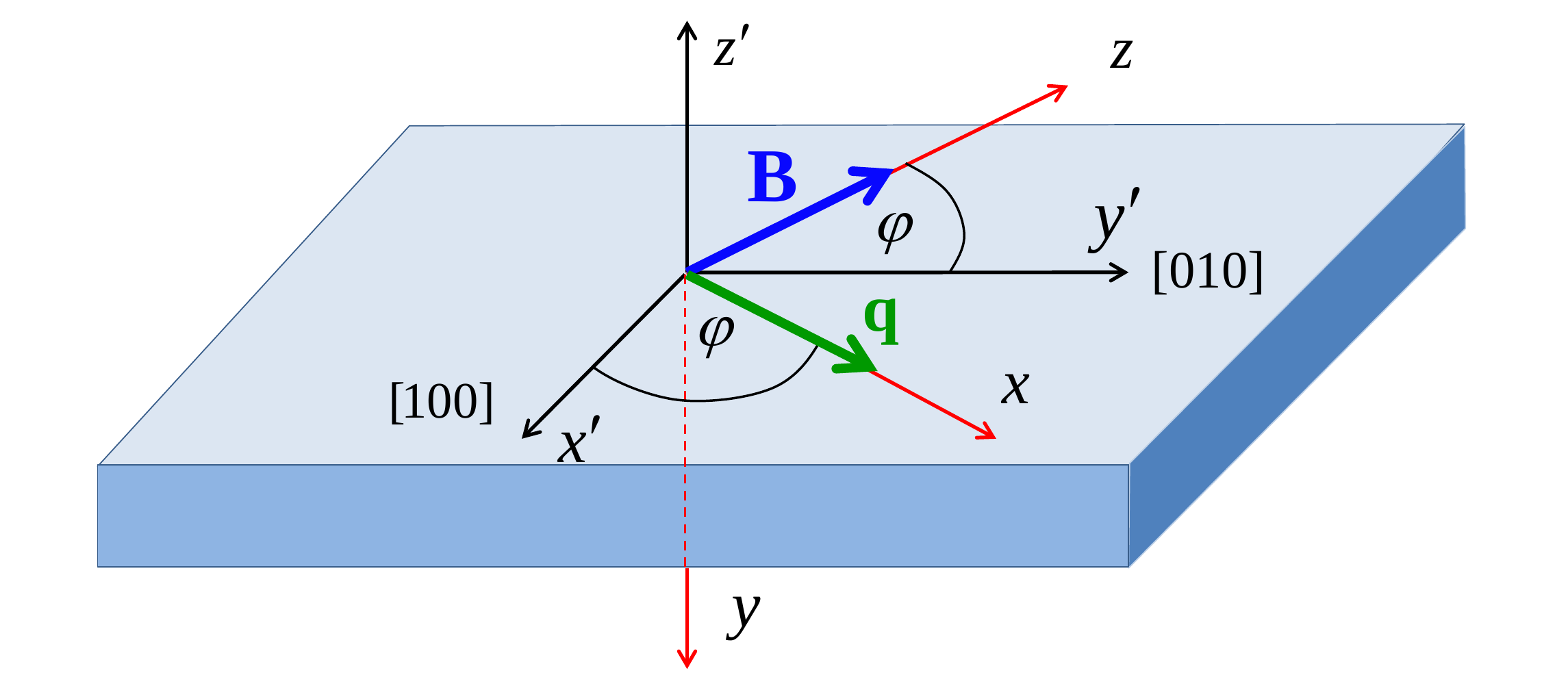}
\caption{(Color online)
Reference frames ${\cal R}'$ (black) and $\cal R$ (red) used to describe the electronic states in a quantum well
with in-plane magnetic field $\bfB$ and
spin-wave propagation direction $\bfq$.} \label{fig1}
\end{figure}

%--------------------------------------------------------------------------------------------------------------------

\section{Quantum well with in-plane magnetic field and SOC} \label{sec:III}

In this Section we will consider the electronic ground state of an $n$-doped semiconductor quantum well
with in-plane magnetic field and Rashba and Dresselhaus SOC, using DFT and the effective-mass approximation.
The setup is illustrated in Figure \ref{fig1},
which defines two reference frames. The reference frame $\cal R'$ is fixed with respect to the quantum well:
the quasi-2DEG lies in the $x' - y'$ plane, where the $x'$-axis points along the crystallographic [100] direction and the $y'$-axis points
along the [010] direction. The $z'$-axis is along the direction of quantum confinement of the well.

The coordinate system $\cal R$ is oriented such that its $x-z$ plane lies in the quantum well plane, and
the $z$-axis points along the in-plane magnetic field $\bfB$. In the inelastic light scattering experiments that
we will discuss below, $\bfB$ is always perpendicular to the wave vector $\bfq$ of the spin waves.
Here, $\bfq$ is along the $x$-axis, which is at an angle $\varphi$ with respect to the $x'$-axis.

The single-particle states in the reference frame $\cal R'$ can be written as
\begin{equation} \label{III.1.1}
\Psi'_{j\bfk}(\bfr') = e^{i\bfk\cdot \bfr'}\psi'_{j\bfk}(z')\:.
\end{equation}
Here, $\bfk = (k_{x'},k_{y'},0)$ is the in-plane wave vector and $j$ is the subband index; in the following,
we are only interested in the lowest spin-split subband, so the subband index $j$ will be replaced by
the index $p=\pm 1$. The two-component spinors
$\psi'_{p\bfk}(z')$ are obtained from the following Kohn-Sham equation:
\begin{equation} \label{III.1.2}
[h_0 \hat \sigma_0 + h_{x'} \hat \sigma_{x'} + h_{y'} \hat \sigma_{y'} ]\psi'_{p\bfk}(z') = E_{p\bfk} \psi'_{p\bfk}(z') \:,
\end{equation}
where $\hat \sigma_0$ is the $2\times 2$ unit matrix.
The spin-independent, diagonal part of the single-particle Hamiltonian is
\begin{equation} \label{III.1.3}
h_0 = \frac{k^2}{2} - \frac{1}{2}\frac{d^2}{d z'^2} + v_{\rm conf}(z') + v_{\rm H}(z') + v_{\rm xc}^{+}(z') \:.
\end{equation}
Here, $v_{\rm conf}(z')$ is the quantum well confining potential (an asymmetric square well), $v_{\rm H}(z')$ is the
Hartree potential, and we define $v_{\rm xc}^{\pm}(z') = [v_{\rm xc \ua}(z') \pm v_{\rm xc \da}(z')]/2$.

The off-diagonal parts in Eq. (\ref{III.1.2}) contain the Zeeman energy $Z$ plus xc and SOC contributions:
\begin{eqnarray}\label{III.1.4}
h_{x'} &=& -\left( \frac{Z}{2}  + v_{\rm xc}^{-}(z')\right) \sin\varphi
+ \alpha k_{y'} + \beta k_{x'} \\
h_{y'} &=& \left( \frac{Z}{2}  + v_{\rm xc}^{-}(z')\right) \cos\varphi
- \alpha k_{x'} - \beta k_{y'}  \:, \label{III.1.5}
\end{eqnarray}
where $\alpha$ and $\beta$ are the standard Rashba and Dresselhaus coupling parameters.

To find the solutions of the Kohn-Sham system, it is convenient to transform  into the reference system $\cal R$ of Fig. \ref{fig1},
whose $z$-axis is along the magnetic field direction. We introduce two in-plane vectors, $\bfq_0$ and $\bfq_1$, whose components
(in the frame ${\cal R}'$) are
\begin{eqnarray} \label{III.1.6}
q_{0 x'} &=& 2(\alpha \cos\varphi + \beta\sin\varphi)\\
q_{0 y'} &=& 2(\alpha \sin\varphi + \beta \cos\varphi) \label{III.1.7}
\end{eqnarray}
and
\begin{eqnarray} \label{III.1.8}
q_{1 x'} &=& 2(-\alpha \sin\varphi + \beta\cos\varphi)\\
q_{1 y'} &=& 2(\alpha \cos\varphi - \beta \sin\varphi) \:. \label{III.1.9}
\end{eqnarray}
With this, Eq. (\ref{III.1.2}) transforms into
\begin{equation} \label{III.1.10}
\left[h_0 \hat \sigma_0 + \left(\frac{Z - \bfk\cdot\bfq_0}{2} +v_{\rm xc}^{-} \right)  \hat \sigma_{z}
+ \frac{\bfk\cdot \bfq_1}{2} \hat \sigma_{x} \right]\psi_{p\bfk} = E_{p\bfk} \psi_{p\bfk}
\end{equation}
(the scalar products $\bfk \cdot \bfq_0$ and $\bfk \cdot \bfq_1$ are invariant under this coordinate transformation).
The solutions of Eq. (\ref{III.1.10}) can be written as follows:
\begin{equation} \label{III.1.11}
E_{p\bfk} = \frac{k^2}{2} + \frac{\varepsilon_{\ua} +\varepsilon_{\da}}{2}
+
\frac{p}{2}\sqrt{ \left(Z^* - \bfk \cdot \bfq_0\right)^2 + (\bfk\cdot \bfq_1)^2 } \:,
\end{equation}
where $Z^* = \varepsilon_{\ua} -\varepsilon_{\da}$ and $p=\pm 1$.
The associated eigenfunctions are
\begin{eqnarray}\label{III.1.12}
\psi_{+,\bfk}(y) &=& \frac{1}{\sqrt{1+b^2}}\left( \begin{array}{c} 1 \\ b \end{array} \right) \phi(y)\\
\psi_{-,\bfk}(y) &=& \frac{1}{\sqrt{1+b^2}}\left( \begin{array}{c} -b \\ 1 \end{array} \right) \phi(y) \label{III.1.13}
\end{eqnarray}
and
\begin{equation}\label{III.1.14}
b = \frac{1}{\bfk \cdot \bfq_1}  \left[
\sqrt{ \left(Z^* - \bfk \cdot \bfq_0\right)^2 + (\bfk\cdot \bfq_1)^2 }
- Z^* + \bfk \cdot \bfq_0  \right].
\end{equation}
The solutions (\ref{III.1.11})--(\ref{III.1.14}) have been expressed in terms of the solutions in the absence of SOC,
$\varepsilon_{\ua,\da}$ and $\phi(y)$, which follow from
\begin{equation} \label{III.1.15}
\left[h_0 \pm \left(\frac{Z}{2} +v_{\rm xc}^{-}\right)\right]\phi_{\ua,\da} = \varepsilon_{\ua,\da}\phi_{\ua\da} \:.
\end{equation}
The spin-up and spin-down envelope functions $\phi_\ua$ and  $\phi_\da$ are practically identical
for the systems considered here, which allowed us to use $\phi_\ua \approx \phi_\da \equiv \phi$ to express the solutions
(\ref{III.1.12}) and (\ref{III.1.13}) in a relatively compact form.

Finally, let us expand the solutions (\ref{III.1.11})--(\ref{III.1.14}) in powers of the SOC coefficients $\alpha$ and $\beta$.
We obtain to second order in SOC
\begin{equation} \label{III.1.16}
E_{p\bfk} = \frac{k^2}{2} + \frac{\varepsilon_{\ua} +\varepsilon_{\da}}{2}
+ \frac{ p }{2} \! \left(Z^* -  \bfk\cdot \bfq_0
+  \frac{ (\bfk\cdot \bfq_1)^2}{2Z^*} \right)
\end{equation}
and
\begin{eqnarray}\label{III.1.17}
\psi_+(y) &=&  \left( \begin{array}{c} \dy
1 - \frac{(\bfk\cdot \bfq_1)^2}{8 Z^{*2}}  \\[2mm] \dy
\frac{\bfk\cdot \bfq_1}{2 Z^*} + \frac{ (\bfk\cdot\bfq_0)(\bfk\cdot \bfq_1) }{2Z^{*2}}\end{array}\right) \! \phi(y) \\
\psi_-(y) &=&  \left( \begin{array}{c} \dy
-\frac{\bfk\cdot \bfq_1}{2Z^*} - \frac{ (\bfk\cdot\bfq_0)(\bfk\cdot \bfq_1)  }{2Z^{*2}}\\[2mm] \dy
1 - \frac{(\bfk\cdot \bfq_1)^2 }{8Z^{*2}} \end{array}\right) \! \phi(y).
\label{III.1.18}
\end{eqnarray}

\begin{figure}
\includegraphics[width=\linewidth]{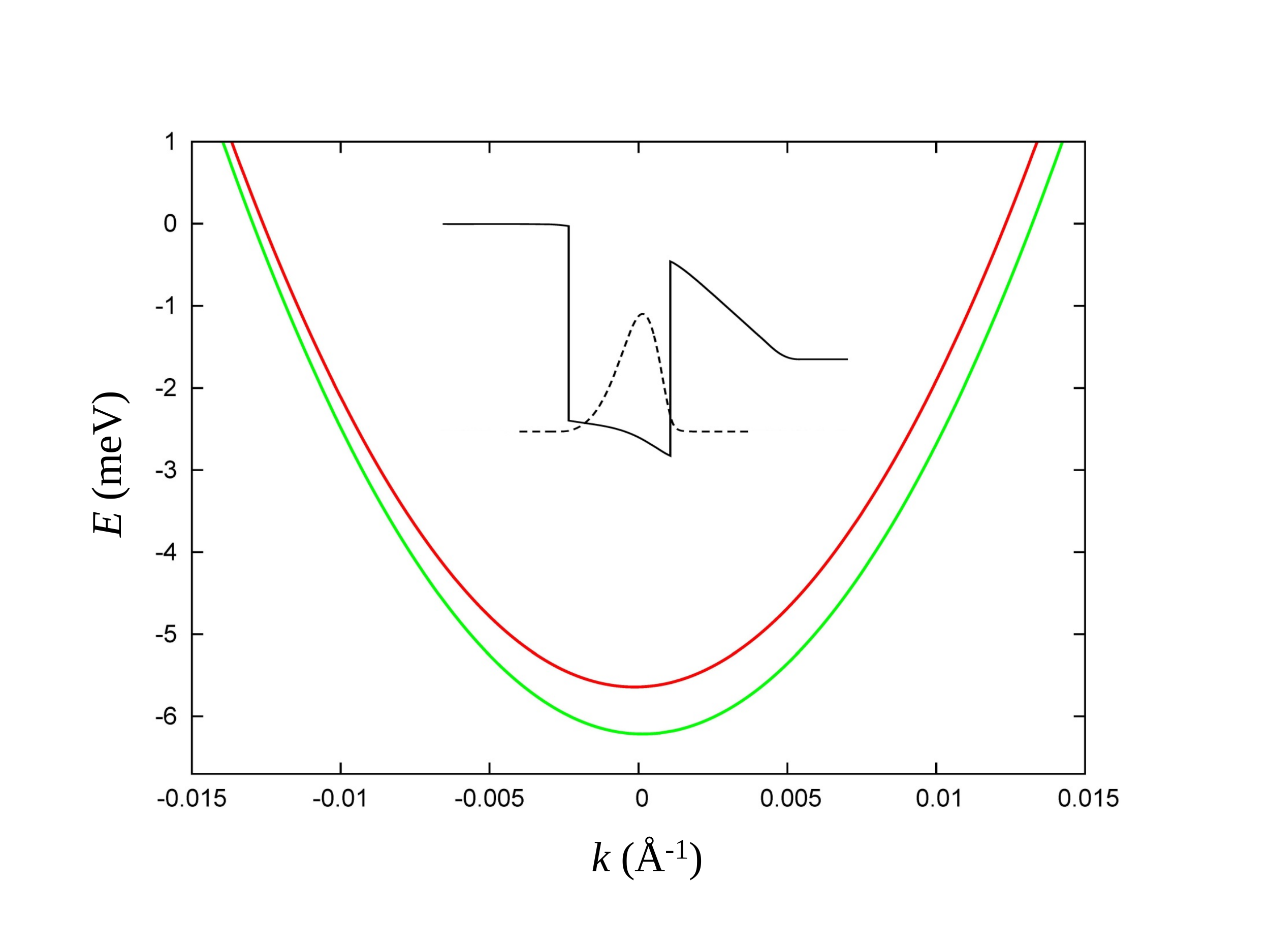}
\caption{(Color online)
Spin-split lowest subband, Eq.  (\ref{III.1.11}), of an asymmetrically doped 20 nm CdTe quantum well with $B=4.18$ T,
with $\alpha=2.2$ meV{\AA} and $\beta = 3.9$ meV\AA, taken at an angle $\varphi=45^{\rm o}$ (i.e. along [110]).
The inset shows the quantum well profile and the electronic density distribution.} \label{fig2}
\end{figure}

We illustrate the energy dispersion (\ref{III.1.11}) of the lowest spin-split subband in Fig. \ref{fig2}.
Here, we consider an asymmetrically doped CdMn quantum well of width 20 nm and electron density $2.6\times 10^{11}$ $\rm cm ^{-1}$.
An applied magnetic field of $B=4.18$ T leads to the bare and renormalized Zeeman energies $Z=0.40$ meV and $Z^*=0.573$ meV,
respectively, using the LDA. Here, we use the effective-mass parameters $m^*=0.105 m$, $e^*=1/\sqrt{10}$, and $g^*=-1.64$ for CdTe.

We choose the Rashba and Dresselhaus parameters $\alpha=2.2$ meV{\AA} and $\beta = 3.9$ meV{\AA} (see below), which causes
the two subband to be slightly displaced horizontally with respect to one another (in Fig. \ref{fig2}, we plot $k$ along
the [110] direction, i.e., for $\varphi = 45^{\rm o}$).

%--------------------------------------------------------------------------------------------------------------------

\section{Spin-flip waves dispersion} \label{sec:IV}

\subsection{Linear-response formalism}
In the following, we are interested in the collective spin-flip modes in a quantum well with in-plane magnetic field and SOC.
Based on the translational symmetry in the $x-z$ plane, one can Fourier transform with respect to
the in-plane position vector $\bfr = (x,z)$; this introduces the in-plane wave vector $\bfq$. The TDDFT linear-response  equation (\ref{2.9}) then becomes
\begin{equation}\label{IV.1}
 n^{(1)}_{\sigma\sigma'}(\bfq,y,\omega)= \sum_{\tau\tau'} \int dy' \chi_{\sigma\sigma',\tau\tau'}(\bfq,y,y',\omega) v^{(1)\rm eff}_{\tau\tau'}(\bfq,y',\omega),
\end{equation}
where the noninteracting response function is given by
\begin{eqnarray} \label{IV.2}
&&\chi_{\sigma\sigma',\tau\tau'}(\bfq,y,y',\omega)
=\nonumber\\
&&-\sum_{pp'}^{\pm 1}  \int\!\frac{d^2k}{(2 \pi)^2}  \:
\frac{\theta(E_F - E_{p \bfk}) }
{\omega - E_{p \bfk} + E_{p' \bfk-\bfq}+ i\eta}\nonumber\\
&&
{}\times \psi_{p \sigma}(\bfk,y) \psi^*_{p' \sigma'}(\bfk-\bfq,y)
\psi^*_{p  \tau}(\bfk,y') \psi_{p' \tau'}(\bfk-\bfq,y') \nonumber\\
&&{}+
\sum_{pp'}^{\pm1}  \int\! \frac{d^2k}{(2 \pi)^2} \:
\frac{\theta(E_F - E_{p \bfk}) }
{\omega + E_{p \bfk} - E_{p' \bfk+\bfq}+ i\eta} \nonumber\\
&&
{}\times \psi_{ p' \sigma}(\bfk+\bfq,y) \psi^*_{p \sigma'}(\bfk,y)
\psi^*_{p' \tau}(\bfk+\bfq,y') \psi_{p \tau'}(\bfk,y') . \nonumber\\
\end{eqnarray}
The energy eigenvalues $E_{p\bfk}$ and the single-particle states $\psi_{p\sigma}(\bfk,y)$
are defined in Eqs. (\ref{III.1.16})--(\ref{III.1.18}). $\theta$ is the step function,
and the Fermi energy is given by $E_F = \pi N_s - (\alpha^2 + \beta^2)$, where $N_s$ is the electronic
sheet density (the number of electrons per unit area). We assume here that both spin-split subbands are occupied, which is
different from the situation considered in Refs. \onlinecite{Ashrafi2012,Maiti2015a,Maiti2015b}.

In the response function (\ref{IV.2})
we only consider spin-flip excitations within the lowest spin-split subband of the quantum well; contributions from higher subbands
are ignored, because they will be irrelevant as long as the Zeeman splitting
is small compared to the separation between the lowest and higher subbands, which is safely the case here.

An interesting property of the response equation (\ref{IV.1}) is that it is invariant under the simultaneous
sign changes $\alpha \to -\alpha$, $\beta \to -\beta$, and $\bfq \to -\bfq$, as can easily be seen from the form of the
response function (\ref{IV.2}). From this we conclude that an expansion of the coefficients $E_0$ and $E_2$
in Eq. (\ref{I.2}) only has even orders of $\alpha,\beta$, while only odd orders of $\alpha,\beta$ contribute
to $E_1$.

The $4\times 4$ matrix response equation (\ref{IV.1}) can be solved numerically, within the ALDA, to yield the spin-wave
dispersions. \cite{Ullrich2002,Ullrich2003} However, much physical insight can be gained by
an analytic treatment, which can be done for small wave vectors $\bfq$: the spin-wave dispersion then takes on the form of Eq. (\ref{I.2}),
and our goal is to determine the coefficients $E_0$  and $E_2$ and compare them to experiment.
We have done this analytically for $E_0$  and numerically for $E_2$, as discussed below.

Instead of the spin-density-matrix response (\ref{IV.1}), it is convenient to
work with the density-magnetization response: we replace the spin-density matrix $n_{\sigma\sigma'}$, defined in Eq. (\ref{2.7}), with the total
density $n\equiv m_0$ and the three components of the magnetization $m_{x,y,z}$ as basic variables. In the following, we
replace the labels $(x,y,z)$ with $(1,2,3)$ to streamline the notation.

The connection between the two sets of variables is made via the Pauli matrices:
\begin{equation}
 m_i(\bfr)=\mbox{tr}\{ \hat \sigma_i \underline{n}(\bfr) \} \:, \qquad i=0,\ldots,3.
\end{equation}
We can also express this through a $4\times4$ transformation matrix $\underline{\underline{T}}$, connecting
the elements $m_i$ and $n_{\sigma\sigma'}$ arranged as column vectors: $\vec{m} = \underline{\underline{T}} \vec{n}$. In detail,
\begin{equation}\label{11}
\left( \begin{array}{c} m_0 \\ m_1 \\ m_2 \\ m_3 \end{array} \right)
=
\left( \begin{array}{cccc}
1 & 0 & 0 & 1 \\
0 & 1 & 1 & 0 \\
0 & i & -i & 0 \\
1 & 0 & 0 & -1
\end{array} \right)
\left( \begin{array}{c} n_{\ua\ua} \\ n_{\ua\da} \\ n_{\da\ua} \\ n_{\da\da} \end{array} \right).
\end{equation}
In a similar way,  one can transform the spin-density-matrix response equation (\ref{IV.1}) into the response equation for the density-magnetization:
\begin{equation}\label{IIII.2}
 m^{(1)}_i(\bfq,y,\omega)= \sum_{k=0}^3\int dy' \Pi_{ik}(\bfq,y,y',\omega) V^{(1)}_k(\bfq,y',\omega)\:,
\end{equation}
where $\underline{\underline{\Pi}}=2\underline{\underline{T}}\, \underline{\underline{\chi}}\, \underline{\underline{T}}^{-1}$ is the
noninteracting density-magnetization response function,
and $\vec{V}^{(1)}=\frac{1}{2}\underline{\underline{T}}\vec{v}^{(1)\rm eff}$ is the effective perturbing potential.

We are only interested in the spin-flip excitations, which are eigenmodes of the system: hence, no external perturbation is
necessary. Furthermore, the Hartree contributions drop out in the spin channel, so the effective potential only consists of the xc part:
\begin{equation}
V^{(1)}_k(\bfq,y,\omega)= \sum^3_{l=0} \int dy' h^{\rm xc}_{kl}(\bfq,y,y',\omega) m^{(1)}_l(\bfq,y',\omega)\:.
\end{equation}
In the ALDA, the xc kernels $h^{\rm xc}_{kl}$ do not depend on frequency and wave vector.\cite{Ullrich2002}
Once we have the density-magnetization response, we can multiply it with the xc matrix.
The xc matrix has a simple form, because in this reference frame the spin polarization direction is along $z$:
\begin{equation}\label{4.1}
\underline{\underline H}^{\rm xc} =
\left( \begin{array}{cccc}
h^{\rm xc}_{00} & 0 & 0 & h^{\rm xc}_{03}\\
0 & h^{\rm xc}_{11} & 0 & 0\\
0 & 0 & h^{\rm xc}_{22} & 0 \\
h^{\rm xc}_{30}&  0 & 0 & h^{\rm xc}_{33}
\end{array}\right)
\end{equation}
where
\begin{eqnarray}
\hxc_{00} &=& 2 \frac{\partial \exc}{\partial n} + n \frac{\partial^2 \exc}{\partial n^2}
- 2\zeta \frac{\partial^2 \exc}{\partial n \partial \zeta} + \frac{\zeta^2}{n}\frac{\partial^2 \exc}{\partial n^2}
\\
\hxc_{03} &=& \hxc_{30} =
\frac{\partial^2 \exc}{\partial n \partial \zeta} - \frac{\zeta}{n} \frac{\partial^2 \exc}{\partial \zeta^2}
\\
\hxc_{11} &=& \hxc_{22} = \frac{1}{n\zeta} \frac{\partial \exc}{\partial \zeta}\\
\hxc_{33} &=&  \frac{1}{n} \frac{\partial^2 \exc}{\partial \zeta^2} \:.
\end{eqnarray}
All quantities are evaluated at the local density $n(y)$ and spin polarization $\zeta(y)$ and multiplied with $\delta(y-y')$.
Here, $e_{\rm xc}$ is the xc energy per particle of the 3D electron gas.\cite{Perdew1992}

To find the collective modes, we can recast the response equation (\ref{IIII.2}) in such a way that the $y$-dependence goes away; the xc kernels
$\hxc_{kl}$ are then replaced by their averages over $\phi^4(y)$. We need to determine those frequencies where the matrix
\begin{eqnarray}\label{4.2}
\underline{\underline M}(\bfq,\omega) = \underline{\underline H}^{\rm xc}(\bfq,\omega) \underline{\underline{\Pi}}(\bfq,\omega)
\end{eqnarray}
has the eigenvalue 1. In other words, we solve the $4\times 4$ eigenvalue problem
\begin{equation} \label{59}
\underline{\underline M}(\bfq,\omega) \vec x = \lambda(\bfq,\omega) \vec x
\end{equation}
and find the mode frequencies by solving $\lambda(\bfq,\omega)=1$ for $\omega$, where $\bfq$ is fixed. In general there will be 4 solutions.
This is in principle exact, provided we know the exact Hxc matrix, which, in general, depends on $(\bfq,\omega)$.
In ALDA, it is a constant (for given density and spin polarization).

\subsection{Beyond Larmor's theorem: leading SOC corrections}

In the presence of SOC, the spin-wave dispersions are modified in an interesting and subtle
manner. For small values of $\bfq$, the spin-wave dispersion has the quadratic form given in Eq. (\ref{I.2}).
Our goal is now to obtain the coefficient $E_0$ to leading order in the Rashba and
Dresselhaus coupling strengths $\alpha$ and $\beta$.
To do this, we carry out a perturbative expansion of the eigenvalue problem (\ref{59})
in orders of SOC. At $q=0$, the matrix can be written as
\begin{equation}
\underline{\underline M}(0,\omega) =
\underline{\underline M}^{(0)} + \underline{\underline M}^{(2)} + \ldots \
\end{equation}
where superscripts indicate the order of SOC (the linear order vanishes at $q=0$).

We first solve the zero-order eigenvalue problem $\underline{\underline M}^{(0)}  \vec x^{(0)} = \lambda^{(0)} \vec x^{(0)}$.
The zero-order spin-flip response function is
\begin{equation}
\underline{\underline{\Pi}}^{(0)}(0,y,y',\omega) = \frac{Z^* \phi^2(y)\phi^2(y')}{\pi (\omega^2 - {Z^*}^2)}\left(\begin{array}{cccc}
0 & 0 & 0 & 0\\
0 & Z^* & - i \omega & 0\\
0 & i \omega & Z^* & 0 \\
0 & 0 & 0 & 0 \end{array}\right) \:.
\end{equation}
Defining
\begin{equation}\label{4.3}
f_T = \int dy \frac{ \phi^4(y)}{\pi n(y) \zeta(y) } \left. \frac{\partial e_{\rm xc}}{\partial \zeta}\right|_{n(y),\zeta(y)} \:,
\end{equation}
where $f_T < 0$, we obtain
\begin{equation}\label{4.4}
\underline{\underline{M}}^{(0)} =\frac{Z^* f_T}{\omega^2 - {Z^*}^2}
\left(\begin{array}{cccc}
0 & 0 & 0 & 0\\
0 & Z^* & - i \omega & 0\\
0 & i \omega & Z^* & 0 \\
0 & 0 & 0 & 0 \end{array}\right).
\end{equation}
This matrix has eigenvalue 1 for
\begin{equation}\label{4.5}
\omega= Z^*+ Z^* f_T= Z
\end{equation}
(we discard the negative-frequency solution) in accordance with Larmor's theorem.
The associated eigenvector is $\vec x^{(0)} = 2^{-1/2}(0,-i,1,0)$.

To obtain the change of the collective spin precession caused by the presence of SOC, we need to determine
$\lambda^{(2)}$. Using perturbation theory we obtain the second-order correction of the eigenvalue as
\begin{equation}
\lambda^{(2)} = [\vec x^{(0)}]^\dagger  \underline{\underline M}^{(2)} \vec x^{(0)}\:,
\end{equation}
To construct $\underline{\underline M}^{(2)}$ we need $\underline{\underline \Pi}^{(2)}(0,\omega) $,
the spin-flip response matrix expanded to  second order in $\alpha$ and $\beta$, which requires a
rather lengthy calculation (see supplemental material\cite{supplemental}). We end up with
\begin{equation}
\lambda^{(2)} =
\frac{2 \pi N_s }{ {Z^*}^2f_T^2 }
\big[ (\alpha^2 + \beta^2)(3f_T+2)+ 2\alpha\beta\sin(2\varphi) (f_T+2)\big]
\end{equation}
The condition $ 1= \lambda^{(0)}+\lambda^{(2)}$ gives the final result
for $E_0$,  see Eq. (\ref{eq_modulation_E0}).

Let us now turn to the other two coefficients in Eq. (\ref{I.2}).
The leading contribution to the linear coefficient $E_1$ is in first order in $\alpha$ and $\beta$,
%and it does not contain contain contributions of higher order;
see Eq. (\ref{eq_modulation_E1}), and was already obtained in Ref. \onlinecite{Perez2016}.
The quadratic coefficient $E_2$ describes the renormalization of the spin-wave stiffness $S_{\rm sw}$ due to SOC.
We did not attempt to derive an analytical expression for $E_2$, as it was done without SOC in Eq. (\ref{2.24}), although this could in principle (and with much effort) be done along the same
lines as for $E_0$. Instead, we extract $E_2$ from a fully numerical solution of the linear-response equation for the spin waves,
which includes all orders of $\alpha$ and $\beta$.

\begin{figure}
\includegraphics[width=\linewidth]{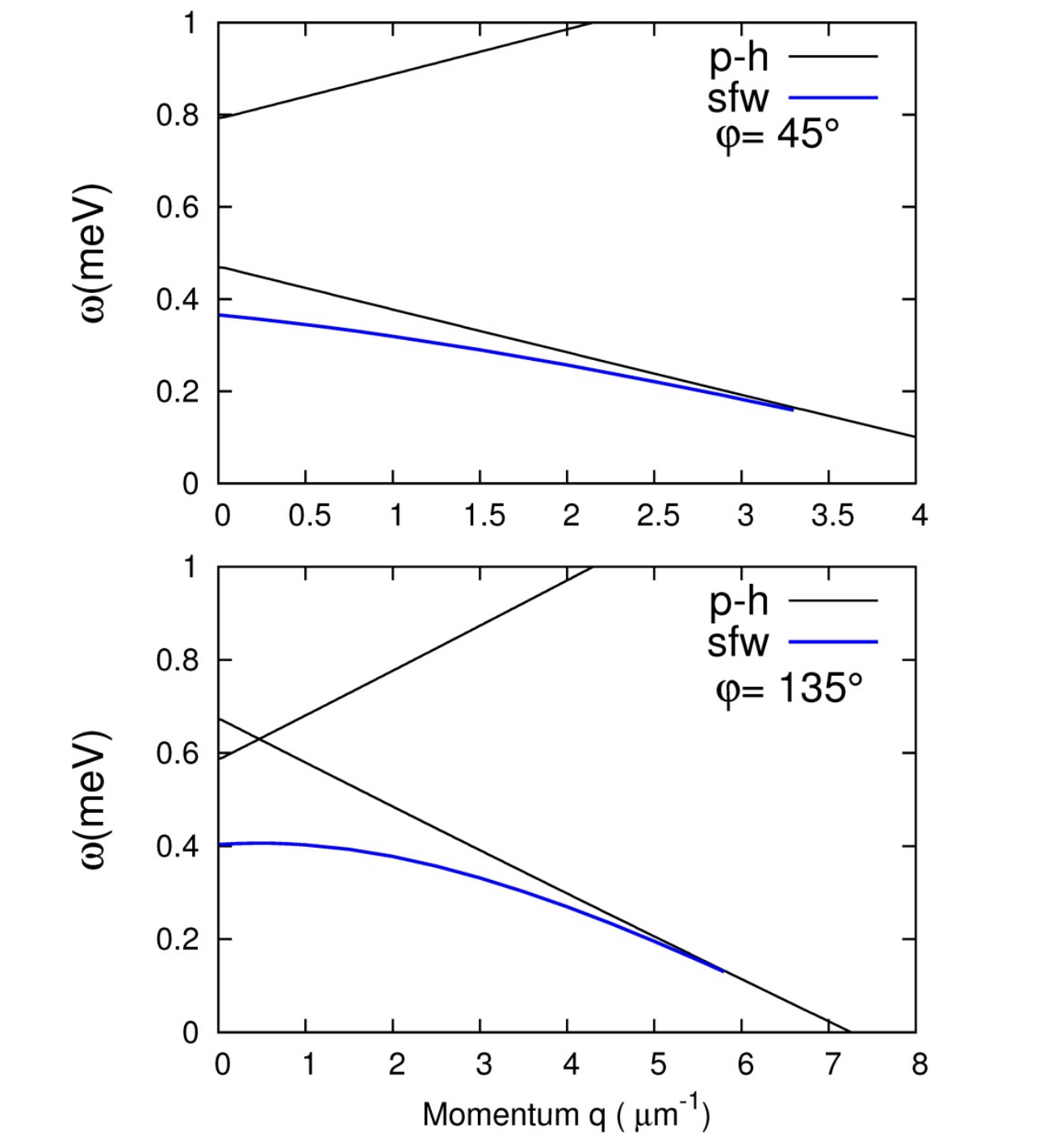}
\caption{(Color online) Spin-flip excitation spectra with SOC for $\varphi=45^{\rm o}$  and $\varphi=135^{\rm o}$, calculated using the ALDA
for the same quantum well as in Fig. \ref{fig2}.
Solid black lines:  boundaries of the single-particle spin-flip continuum. Blue dashed lines:
spin-wave dispersions.}  \label{fig3}
\end{figure}

\section{Results and discussion} \label{sec:V}

According to the theory presented above, the spin-flip excitations in a 2DEG in the presence of SOC
depend on the direction of the applied magnetic field (direction $z$ in Fig. \ref{fig1}).
Figure \ref{fig3} depicts the spin-excitation spectra
for $\varphi=45^{\rm o}$  and $\varphi=135^{\rm o}$, calculated using ALDA, for the same quantum well as in Fig. \ref{fig2}.
Clearly, the spin-wave dispersions and single-particle spin-flip continua differ drastically,
depending on the direction of the in-plane momentum.  In the following, we will compare our theory with experiment.

\subsection{Electronic Raman scattering}

\begin{figure}
\includegraphics[width=0.9\linewidth]{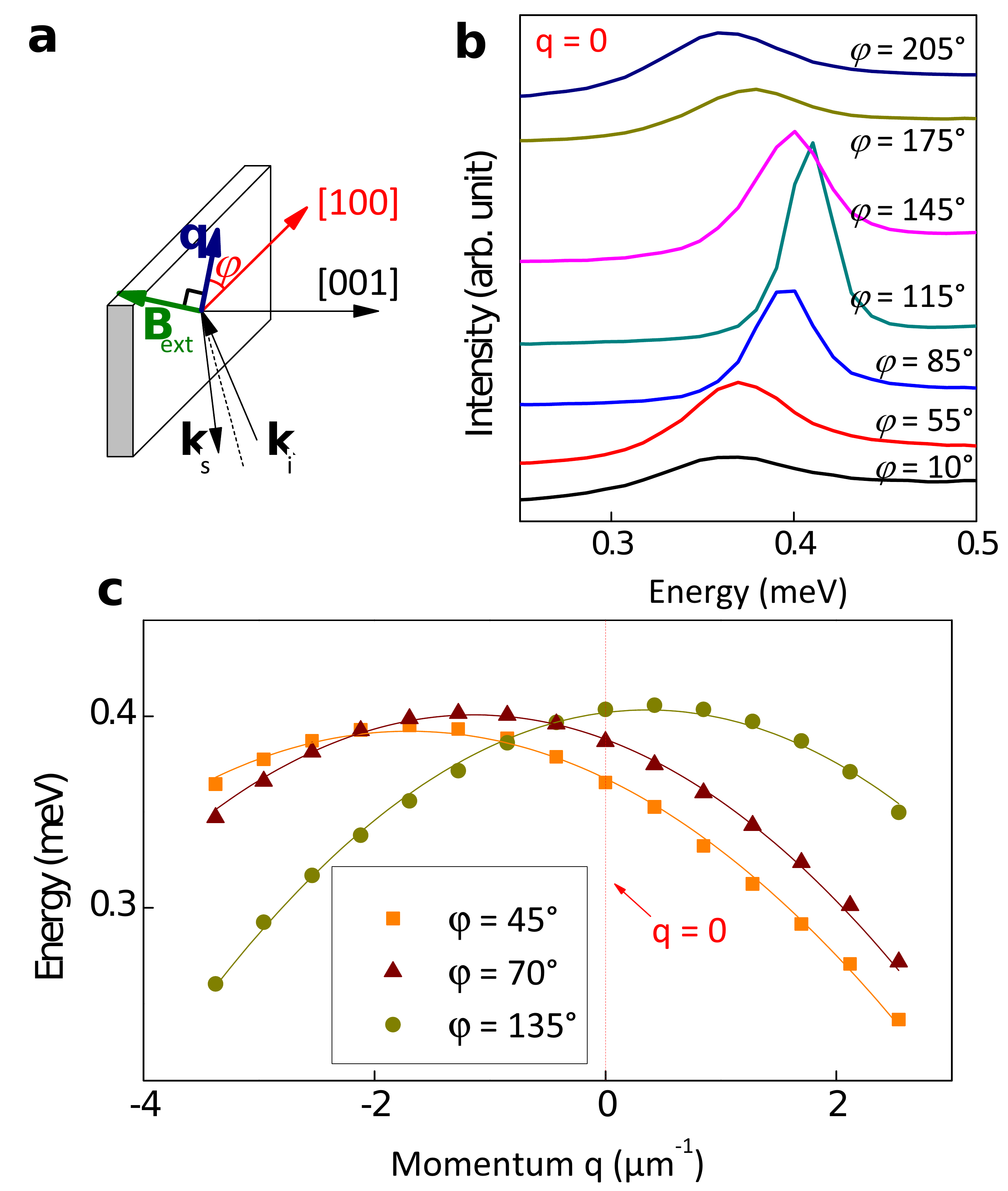}
\caption{(Color online)
(a) Electronic Raman scattering geometry: $\boldsymbol{k}_i$ and $\boldsymbol{k}_s$ are the incoming and scattered light wave vectors, respectively; $\mathbf{q}$ is the transferred
momentum, of in-plane orientation measured by the angle $\varphi$ from $[100]$.
An external magnetic field $\mathbf{B}_{\mathrm{ext}}$ is applied perpendicularly to $\mathbf{q}$.
(b) Raman spectra of the spin wave, obtained at $B_{\mathrm{ext}}=2$ T and $q=0$, for a series of in-plane angles $\varphi$.
(c) Momentum dispersion of the spin wave for different in-plane angles.} \label{fig4}
\end{figure}

We use electronic Raman scattering, whereby a well-controlled in-plane momentum ${\bf q}$ is transferred to the spin excitations of the 2DEG. Under the quasi-scattering geometry shown in Fig. \ref{fig4}a, the transferred momentum is given by
${\bf q}=\boldsymbol{\kappa}_{i,\parallel}-{\boldsymbol{\kappa}}_{s,\parallel} \simeq \frac{4\pi}{\lambda}  \sin \theta \,\boldsymbol{e}_x$, where $\boldsymbol{\kappa}_{i}$ and $\boldsymbol{\kappa}_{s}$ are the wave vectors of the linearly cross-polarized incoming and scattered photons, and $\lambda$ is the incoming wavelength. Our setup allows us to vary ${\bf q}$ both in magnitude and in-plane orientation,
while the magnetic field $\mathbf{B}_{\mathrm{ext}}$ is applied in the plane of the well and always perpendicular to $\mathbf{q}$.

Our sample is an asymmetrically modulation-doped, $20$ nm-thick Cd$_{1-x}$Mn$_{x}$Te ($x \simeq 0.13 \%$) quantum well, grown along the $\left[001\right] $ direction by molecular beam epitaxy, and immersed in a superfluid helium bath  at temperature 2 K. The density of the electron gas is $N_s=2.6\times 10^{11}$ cm$^{-2}$ and the mobility is $1.7 \times 10^5$ cm$^{2}$V$^{-1}$s$^{-1}$.
The small concentration of Mn introduces localized magnetic moments into the quantum well, which are polarized by the external
$B$-field, and act to amplify it.\cite{Perez2007,footnote}

Figure \ref{fig4}b shows a series of spin-wave Raman lines obtained at fixed $B_{\mathrm{ext}}=2$ T and $q=0$, and for various in-plane angles $\varphi$. We observe a clear modulation of the spin-wave energy with  $\varphi$, evidencing the above predicted breakdown of Larmor's theorem.

To better understand the phenomenon, we measure the full spin-wave dispersion by varying the transferred momentum $q$.
Fig. \ref{fig4}c shows the dispersions for three different values of $\varphi$:
they exhibit a quadratic dependence with $q$, with a maximum shifted from the zone center. This shift from the zone center is well understood in the frame of the spin-orbit twist model: \cite{Perez2016} SOC produces a rigid shift of the spin-wave dispersion by a momentum $-{\bf q_0}$,
 see Eq. (\ref{III.1.6}), which depends on $\varphi$. This produces  the linear term  in $q$ in the energy dispersion  of Eq. (\ref{I.2}).

We have systematically measured the spin-wave dispersions for angles $\varphi$ between zero and $360^{\rm o}$;
for each angle, the data are fit to a parabola (as in Fig. \ref{fig4}c), which allows us to extract the coefficients $E_{0,1,2}(\varphi)$.
The experimental results are shown in both Figs. \ref{fig5} and \ref{fig6} (dots),
clearly exhibiting the predicted sinusoidal modulations.

The modulation of $E_0$, with a relative amplitude of about 6\%, demonstrates the breakdown of Larmor's theorem. This effect is
of second order in the SOC. By contrast, the modulation of $E_1$ is a first-order SOC effect.
Another second-order SOC effect is the modulation of the curvature of the spin-wave dispersion, i.e. the spin-wave stiffness $\sws$.
The bottom panels of Figs. \ref{fig5} and \ref{fig6} show the curvature $E_2=\sws / 2$  as a function the in-plane angle
$\varphi$. Again, a sinusoidal variation is observed, with a relative amplitude of about $10 \%$; the phase of
the modulation is opposite to that of $E_0$ and $E_1$.

\subsection{Comparison with theory}

\begin{figure}
\includegraphics[width=1.0\linewidth]{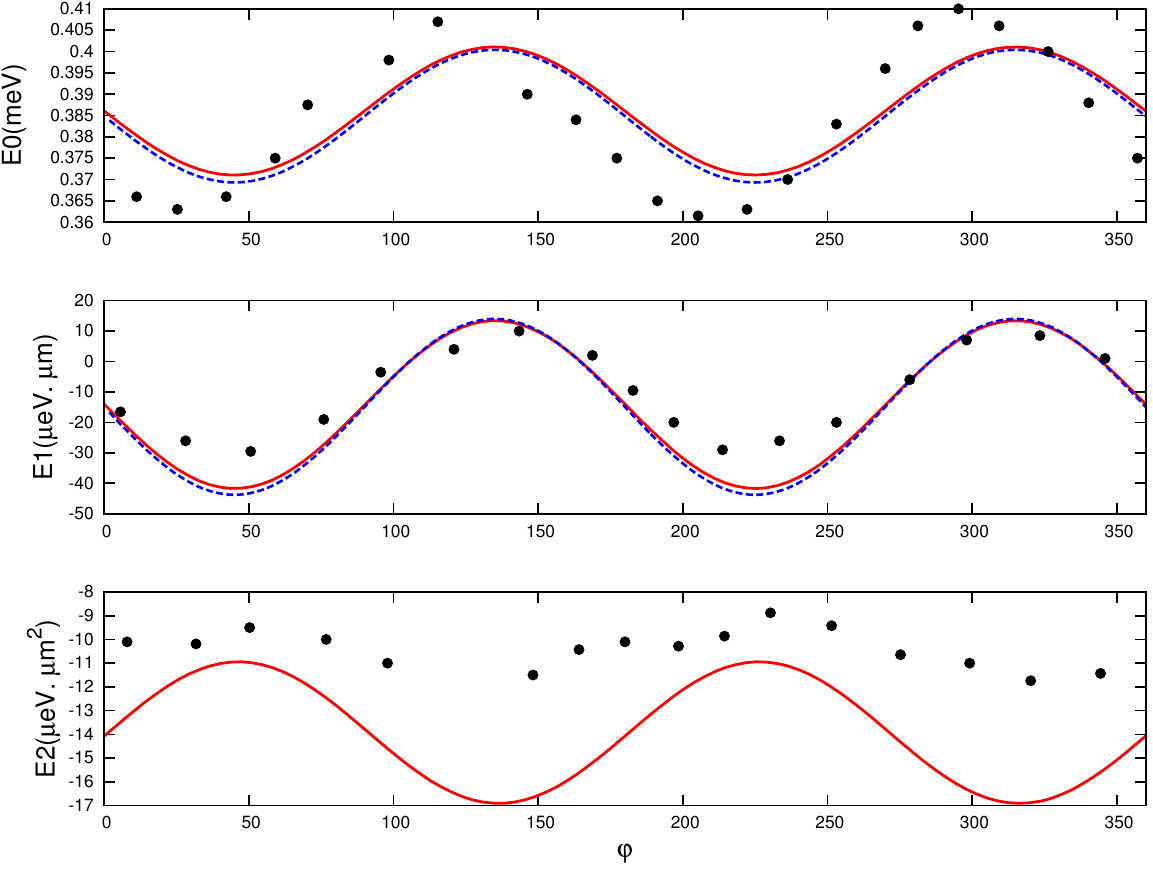}
\caption{(Color online) Coefficients $E_0$, $E_1$, and $E_2$ of the  spin-wave dispersion, Eq. (\ref{I.2}), as a function
of angle $\varphi$. Dots: experimental data. Lines: theoretical results using $Z^*=0.573$ meV obtained with ALDA,
and $\alpha = 1.6$ {meV\AA} and $\beta=3.1$ {meV\AA} obtained by fitting $E_0$ and $E_1$.
The red lines follow from the fully
numerical solution of Eq. (\ref{59}), the dashed blue lines follow from the analytical formulas
(\ref{eq_modulation_E0}) and (\ref{eq_modulation_E1}).}  \label{fig5}
\end{figure}

\begin{figure}
\includegraphics[width=1.0\linewidth]{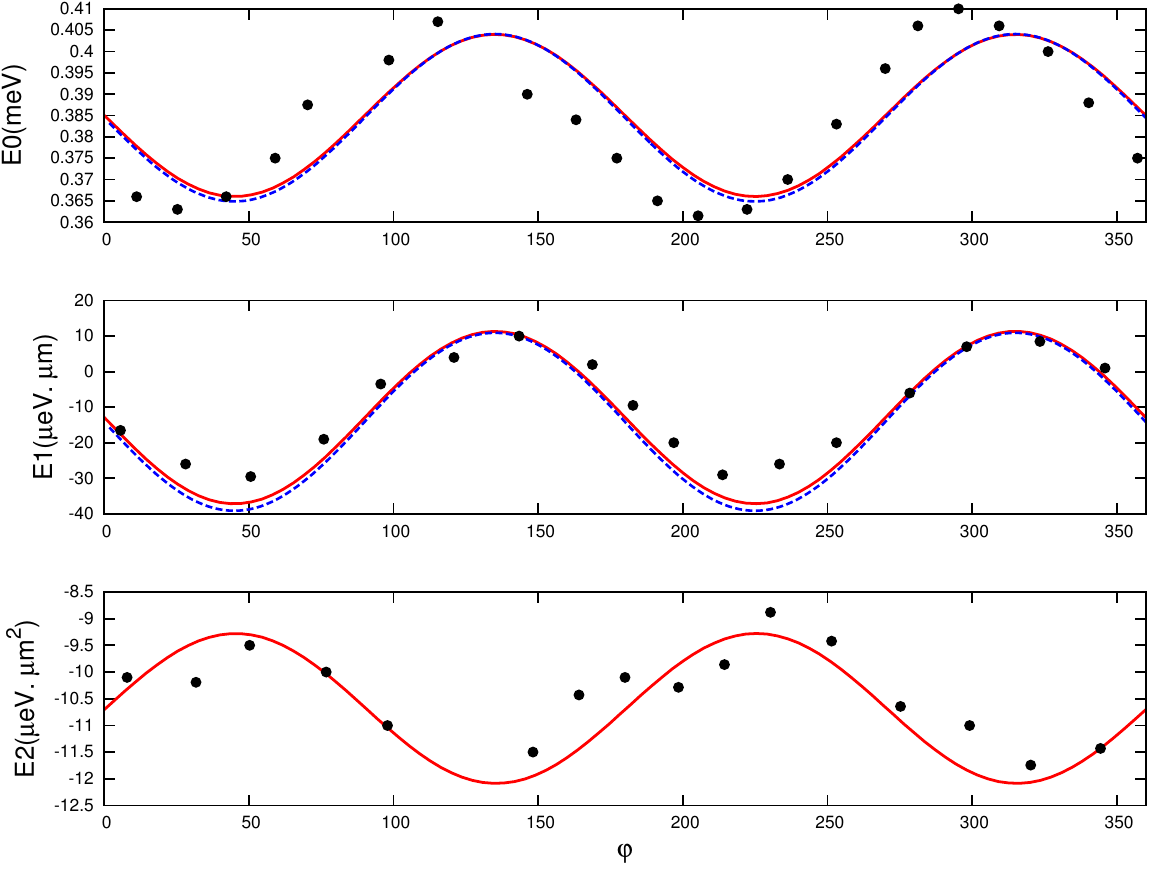}
\caption{(Color online) Same as Fig. \ref{fig5}, but using $Z^*=0.63$ meV, $\alpha=2.2$ {meV\AA}, and $\beta=3.9$ {meV\AA} obtained
from a best fit to the experimental data.
} \label{fig6}
\end{figure}

In Figures \ref{fig5} and \ref{fig6}, the experimental data for $E_0(\varphi)$, $E_1(\varphi)$, and $E_2(\varphi)$ is compared
with theory (lines). In our calculations, we consider, as before, a CdTe quantum well of width 20 nm and density
$N_s=2.6\times 10^{11}$ $\rm cm ^{-2}$. The value of bare Zeeman splitting $Z$ is extracted from the data as follows.
According to Eq. (\ref{eq_modulation_E0}), $E_0$ can be written in the form $E_0(\varphi) = Z - a -b \sin(2\varphi)$. For the range of input
parameters $\alpha$, $\beta$, $Z$ and $Z^*$ under consideration (see below),
the ratio $b/a \approx 1.5$ is almost constant. We temporary fix this ratio, and a fit with the data from
the top panel of Fig. \ref{fig5} then yields $Z = 0.40$ meV and $b=0.024$ meV to within about 3  $\mu$eV.
We can then calculate $Z^*$  using the ALDA xc kernel
[see Eq. (\ref{4.5})], where  $Z^*_{\rm ALDA} = Z/(1+f_T)= 0.573$ meV.
Now fixing $Z$, $Z^*$ and letting $b/a=1.5$, we fit $\alpha$ and $\beta$ from $E_0(\varphi)$ and $E_1(\varphi)$.
An optimal agreement with the experimental results for $E_0$ and $E_1$ is achieved with
$\alpha = 1.6$ { meV\AA} and $\beta = 3.1$ {meV\AA}.

Having determined the set of parameters $Z$, $Z^*$, $\alpha$ and $\beta$, we run the fully
numerical solution of the linear-response equation (\ref{59})
for the spin-flip waves, and fit the small-$q$ dispersion
to a parabola for a given angle $\varphi$ to extract $E_0$, $E_1$, and $E_2$.
As shown in Fig. \ref{fig5}, both the analytical formulas of Eqs. (\ref{eq_modulation_E1}) and (\ref{eq_modulation_E0})
and the numerical solutions (the dashed blue and solid red lines, respectively) are in
very good agreement with the experimental data for $E_0$
and $E_1$, apart from a shift in the phase of the experimental modulation of $E_0$, which is not accounted for by the
theory. It is likely due to an in-plane anisotropy of the g-factor, \cite{Eldridge2013} neglected in the above analysis.

An additional observation from  Fig. \ref{fig5} is that the
analytical formulas and the numerical results for $E_0$ and
$E_1$ are extremely close to each other. This is not surprising, since the next higher-order corrections to $E_0$ and $E_1$ are of fourth and third order in $\alpha,\beta$, respectively (as we showed in Section IV.A), and hence negligible.

On the other hand, the bottom panel of Fig. \ref{fig5} shows
that the calculation dramatically fails to reproduce $E_2$.
Therefore, we repeated the calculations, but now using
a renormalized Zeeman energy $Z^*$ that does not follow
from the ALDA, but from a numerical fit. We fit the numerical solutions with $Z^*$, $\alpha$ and $\beta$ and then
find that using $\alpha=2.2$ {meV\AA}, $\beta = 3.9$ {meV\AA} and $Z^*_{\rm fit}=0.63$ meV
we obtain an excellent agreement with the experimental results for all three modulation parameters, $E_0$, $E_1$, and $E_2$,
as shown in Fig. \ref{fig6}.

The comparison between theory and experiment of the spin-wave modulation parameters thus demonstrates that
the ALDA underestimates $Z^*$ by about 10\%, which seems to be a relatively minor deviation. However,
$E_0$, $E_1$, and $E_2$ depend very sensitively on $Z^*$, which suggests a need for a more accurate description
of dynamical xc effects beyond the ALDA.

\subsection{Density dependence of $E_0$}

\begin{figure}
\includegraphics[width=0.8\linewidth]{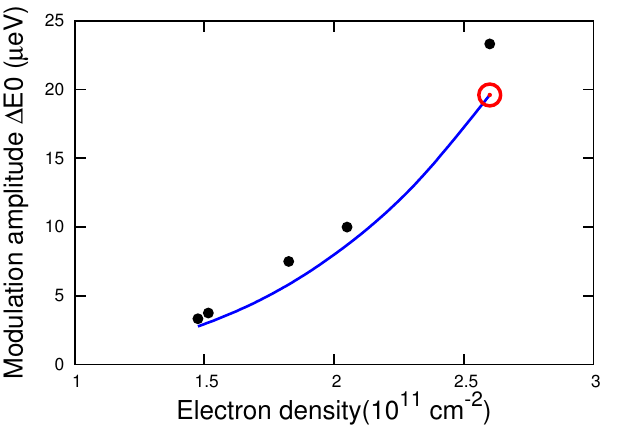}
\caption{(Color online)
Amplitude of the modulation of the $q=0$ spin-wave energy, $\Delta E_0 = (\text{Max} E_0  - \text{Min} E_0)/2$, as a function of the
sheet density $N_s$ of the electron gas in the quantum well. Black dots: experimental data.
Blue line: analytical results using Eq. (\ref{eq_modulation_E0}).}
\label{fig7}
\end{figure}

To further test our theoretical prediction for the breakdown of Larmor's theorem [Eq. \eqref{eq_modulation_E0}],
we will now explore the density dependence of the parameter $E_0$.
In order to vary the electronic density in our sample, we shine  an additional continuous-wave green laser beam ($514.5$ nm) on the quantum well.
This illumination is above the band gap and generates electron-hole pairs in the barrier layer: the electrons neutralize some donor elements of the doping plane, while the holes migrate to the quantum well where they capture free electrons. This leads to a depopulation of the electron gas, which can be precisely controlled by the power of the above-gap illumination.\cite{Baboux2015}
%We calibrate the density changes caused by the green illumination by extracting the Fermi energy from the width of the photoluminescence (PL) spectra, and from the slope of the cross-polarized Raman spectra of the single-particle excitations at zero spin polarization \cite{Baboux2015}.
Using this technique, the density in our sample can be reproducibly reduced by up to a factor $2$.
We measured $E_0(\varphi)$ for different values of $N_s$, and plot
in Fig. \ref{fig7} the amplitude of the $q=0$ modulation (solid circles),
$\Delta E_0 = (\text{Max} E_0  - \text{Min} E_0)/2$, as a function of the electron density.

Again, the data is well reproduced by the analytical result of Eq. \eqref{eq_modulation_E0} (blue line).
The red circle represents the amplitude of $E_0$ for the reference density $N_s^{\rm ref}=2.6\times 10^{11} \: \rm cm^{-2}$, obtained
from our numerical fit in the top panel of Fig. \ref{fig6}. To generate the blue line, we need
$Z^*$ as a function of $N_s$, which we approximate as
\begin{equation}
Z^*(N_s) \approx Z^*_{\rm fit}(N_s^{\rm ref})\frac{ Z^*_{\rm ALDA}(N_s)}{Z^*_{\rm ALDA}(N_s^{\rm ref})}
= 1.10 \, Z^*_{\rm ALDA}(N_s)\:,
\end{equation}
i.e., we approximate the density scaling using the ALDA. We also need the density dependence of the Rashba and Dresselhaus
parameters $\alpha,\beta$. We approximate their density scaling using the $\bfk\cdot {\bf p}$ results
of Ref. \onlinecite{Baboux2015}. Both approximations are well justified by the excellent agreement between theory and experiment
in Fig. \ref{fig7}.

\section{Conclusions}  \label{sec:VI}

In this paper, we presented a detailed theoretical and experimental study of spin-wave dispersions in a 2DEG
in the presence of Rashba and Dresselhaus SOC. In earlier work\cite{Perez2016} we had limited ourselves to the leading
(first-order) SOC effects, which causes a momentum-dependent shift of the spin-wave dispersions, but leaves the spin-wave
stiffness as well as Larmor's theorem intact. We have now discovered some subtle corrections which arise when
second-order SOC effects are taken into account: Larmor's theorem is broken, and
the spin-wave stiffness is modified. Both corrections are relatively small (of order 10\% or less) but experimentally detectable.

We presented a linear-response theory, based on TDDFT, to fully account for SOC effects to first, second and higher orders in SOC.
A detailed comparison with experimental data, obtained using inelastic light scattering, confirmed the accuracy of the theory
and allowed us to extract the SOC parameters $\alpha$ and $\beta$, as well as the renormalized Zeeman splitting $Z^*$.

A major outcome of our study is that we discovered that the ALDA does not lead to a satisfactory description of the second-order
SOC modulation effects of the spin waves.  At present, there are only few approaches in ground-state DFT for noncollinear magnetism
that go beyond the LDA, such as the optimized effective potential (OEP)\cite{Sharma2007} or gradient corrections.\cite{Scalmani2012,Eich2013a,Eich2013b} This provides motivation for the search for better xc functionals in TDDFT
for noncollinear spins. In particular, any such new xc functional should be well-behaved in the crossover between three-and two-dimensional
systems.\cite{Karimi2014}

The study of spin waves in electron gases confined in semiconductor quantum wells under the presence of SOC is also of practical
interest. Manipulation of the Rashba
and Dresselhaus coupling strengths can be used to control the spin-wave group velocity.\cite{Perez2016}
Since spin waves can be used as carriers of spin-based information, this may lead to applications in spintronics.
Here we have provided a suitable theoretical framework to describe these effects.

\begin{acknowledgments}
S.K. and C.A.U. are supported by DOE Grant DE-FG02-05ER46213.
F.B. and F.P. acknowledge support from the Fondation CFM, C'NANO IDF and ANR.
The research in Poland was partially supported by the National Science Centre
(Poland) through grants DEC-2012/06/A/ST3/00247 and DEC-2014/14/M/ST3/00484.
\end{acknowledgments}


\begin{thebibliography}{99}
\bibitem{footnote}
In our numerical calculations, the Mn magnetic moments are not included, which does not
affect the electronic structure and dynamics. However, we need to compensate by using
a larger magnetic field (4.18 T) compared to experiment (2 T).
\end{thebibliography}
\end{document}